\documentclass{aa}  

\usepackage{url}
\usepackage{graphicx}
\usepackage{soul}
\usepackage{float} 
\usepackage{placeins}
\usepackage{txfonts}
\usepackage[dvipsnames]{xcolor}
\usepackage{verbatim}
\usepackage{url}
\usepackage{hyperref}
\begin{document} 

   \title{Are Solar Microflares more efficient in accelerating electrons when rooted near a Sunspot?}

  \author{Jonas Saqri\inst{1} \and
      Astrid M. Veronig\inst{1,2} \and
        Andrea Francesco Battaglia\inst{3,4} \and
      Ewan C. M. Dickson\inst{1} \and
    Dale E. Gary\inst{5}\and
    S\"am Krucker\inst{3,6}
     }

   \institute{
                 Institute of Physics, University of Graz, A-8010 Graz, Austria
             \and
        Kanzelhöhe Observatory for Solar and Environmental Research, University of Graz, A-9521 Treffen, Austria
             \and
             University of Applied Sciences and Arts Northwestern Switzerland, Bahnhofstrasse 6, 5210 Windisch, Switzerland
                          \and
                                       ETH Z\"urich, R\"amistrasse 101, 8092 Z\"urich, Switzerland 
                         \and
             Center for Solar–Terrestrial Research, New Jersey Institute of Technology, Newark, NJ 07102, USA
             \and
             Space Sciences Laboratory, University of California, 7 Gauss Way, 94720 Berkeley, USA
}
\authorrunning{J. Saqri et~al.}

   \date{}
 
  \abstract
   {
    The spectral shape of the X-ray emission in solar flares varies with the event size, with small flares generally exhibiting softer spectra than large events indicative of a relatively lower number of accelerated electrons at higher energies.
}
   {
    We investigate two microflares of GOES classes A9 and C1 (after background subtraction) observed by STIX onboard Solar Orbiter with exceptionally strong nonthermal emission. We complement the hard X-ray imaging and spectral analysis by STIX with co-temporal observations in the (E)UV and visual range by AIA and HMI, in order to investigate what makes these microflares so efficient in high-energy particle acceleration. 
   }
   {
   A preselection of events in the STIX flare catalog was made based on the ratio of the thermal to nonthermal quicklook X-ray emission. STIX spectrogram science data were used to perform spectral fitting to identify the nonthermal and thermal components. STIX X-ray images were reconstructed to analyze the spatial distribution of the precipitating electrons and the hard X-ray emission they produce. EUV images from SDO/AIA and SDO/HMI LOS magnetograms were analyzed to better understand the magnetic environment and the chromospheric and coronal response. For the A9 event, EOVSA microwave observations were available allowing for image reconstruction in the radio domain. 
   }
   {
    We performed case studies of two microflares observed by STIX on October 11, 2021 and November 10, 2022 that showed unusually hard microflare X-ray spectra with 
    power--law indices of the electron flux distributions $\delta = (2.98 \pm 0.25)$ and $\delta = (4.08 \pm 0.23)$ during their nonthermal peaks and photon energies up to 76\,keV and 50\,keV\,respectively. For both events under study, we found that one footpoint is located within a sunspot covering areas with mean magnetic flux densities in excess of 1500\,G, suggesting that the hard electron spectra are caused by the strong magnetic fields in which the flare loops are rooted. In addition, we revisited the unusually hard RHESSI microflare initially published by \cite{Hannah2008MFlare} and found that in this event also one flare kernel was located within a sunspot, which corroborates the result from the two hard STIX microflares under study.
   }
   {
   The characteristics of the strong photospheric magnetic fields inside sunspot umbrae and penumbrae where the flare loops are rooted play an important role in the generation of exceptionally hard X-ray spectra in these microflares.
   }

   \maketitle
\section{Introduction}
Solar flares are the result of the impulsive release of stored magnetic energy in the solar atmosphere that produces enhanced emission across a wide range of the electromagnetic spectrum. Part of the liberated energy is transferred to accelerated electrons, which can be indirectly observed at hard X-ray wavelengths by bremsstrahlung emission when energetic electrons interact with the cooler and denser plasma in lower atmospheric layers. The recorded X-ray spectrum therefore contains information about the energy distribution of nonthermal electrons, which usually are described by a power-law  (\citealt{brown1971,linHudson1976,holman2011}). 
\par
The occurrence rate of flares increases strongly for smaller sizes/energy content, with the flare frequency distribution usually being described by a power law of slope between 1.5--2.5 derived from different studies of flaring events ranging from nanoflares, microflares to the largest flares observed \citep{dennis1985,Crosby1993,benzKrucker2002,Veronig2002,christe2008,hannah2011,purkhart2022}. The term microflare commonly describes events that release energies of the order of $10^{27}$\,erg, about six orders of magnitude less than the largest solar flares \citep{hannah2011}. It is believed that many processes of larger flares also operate during minor events, but may not be observable as they are masked by the background variations. Other aspects such as the highest energy particles that are typically accelerated in an event may depend on the flare size, with exceptions being observed \citep{Hannah2008MFlare, Ishikawa2013, battaglia2023}.  
\par
The nonthermal part of solar flare X-ray spectra (usually observed around above 10\,keV) can often be fitted by a power law: $I(\epsilon) \propto \epsilon ^{-\gamma}$ at photon energy $\epsilon$ with the photon spectral index $\gamma$ \citep{holman2011}. It has been found in statistical studies that smaller events usually show considerably softer X-ray spectra, with a median value of $\gamma = 6.9$ \citep{hannah2008} for microflares compared to a mean value of $\gamma = 3.9$ for larger flares observed above 30\,keV as reported by \cite{bromund1995}. For a thick target source region, this photon power law index $\gamma$ is related to the spectral index of the underlying electron flux distribution $\delta$ via $\gamma = \delta-1$ \citep{brown1971,holman2011}. As the number of small events is much larger than that of big flares, some outliers of small flares with hard spectra have previously been observed in the RHESSI flare dataset. \cite{Ishikawa2013} studied six B-class microflares with joint RHESSI and WAM observations and found photon spectral indices $\gamma$ of the nonthermal spectra between 3.3 and 4.5 and electron energies up to at least 100\,keV. So far, the microflare with the hardest HXR spectrum reported is the exceptional GOES A7 event studied in \cite{Hannah2008MFlare} which showed strong non-thermal emission up to energies of over 50\,keV and a hard photon power law index of $\gamma = 2.4$. 

\par
Microflares have been found to occur exclusively in active regions \citep{stoiser2007,hannah2011_statistics}, preferably near magnetic neutral lines \citep{liu2004}. In general, they do not occur directly within sunspots but in plages \citep{li1998}. For bipolar configurations, loop--like morphologies are observed with two footpoints located at opposite magnetic polarities and a thermal source between them \citep{liu2004, stoiser2007}. Jet--like morphologies are also observed in association with microflares. They often show three HXR footpoints, consistent with the interchange reconnection scenario commonly invoked to explain open field lines responsible for the escaping plasma and particles \citep{krucker2011}. For a recent case study see \cite{battaglia2023}. 
\par
The Spectrometer-Telescope for Imaging X-rays (STIX; \citealt{krucker2020}) onboard the Solar Orbiter spacecraft operates since 2020. Since then it has observed over 20 000 flares. For this study, we searched this extensive STIX flare catalog for microflares with exceptionally hard spectra with the aim to better understand what makes them so efficient at accelerating electrons to high energies.

\section{Data and Methods}
The Solar Orbiter mission is designed to approach the Sun to within 0.28\,AU and to reach heliographic latitudes of up to 30\,degrees over the course of the mission's duration \citep{mueller2020}, enabling observations with increased instrument sensitivities. STIX onboard Solar Orbiter provides X-ray imaging and spectroscopy of the Sun from 4--150\,keV with an energy resolution of 1\,keV at 6\,keV and a temporal resolution of up to 0.1\,seconds \citep{krucker2020}. STIX uses an indirect imaging concept based on 32 individual detectors subdivided into pixels with each detector being located behind fine grids of varying slit width and orientation, resulting in Moire patterns that encode the spatial distribution of the incoming X-ray emission \citep{krucker2020}. 
With its energy sensitivity and imaging capability, it is able to observe the evolution of thermal as well as nonthermal flare emission to study electron acceleration and plasma heating during solar flares.
\par
From the STIX flare catalog, we selected events up to GOES class C1. We used the recorded STIX quicklook (QL) counts as a proxy for the flare class by using the empirical relation between the GOES 1--8\,\AA\, flux $f$ in $\mathrm{W\,m^{-2}}$ and the distance corrected STIX peak 4--10\,keV QL count rate $X^{'}$: $\mathrm{log_{10}}(f)=0.622-7.376\,\mathrm{log_{10}}(X^{'})$ published by \cite{STIX_data_center}.
\par
\begin{figure}[!h]   
	\includegraphics[width=0.5\textwidth,clip=]{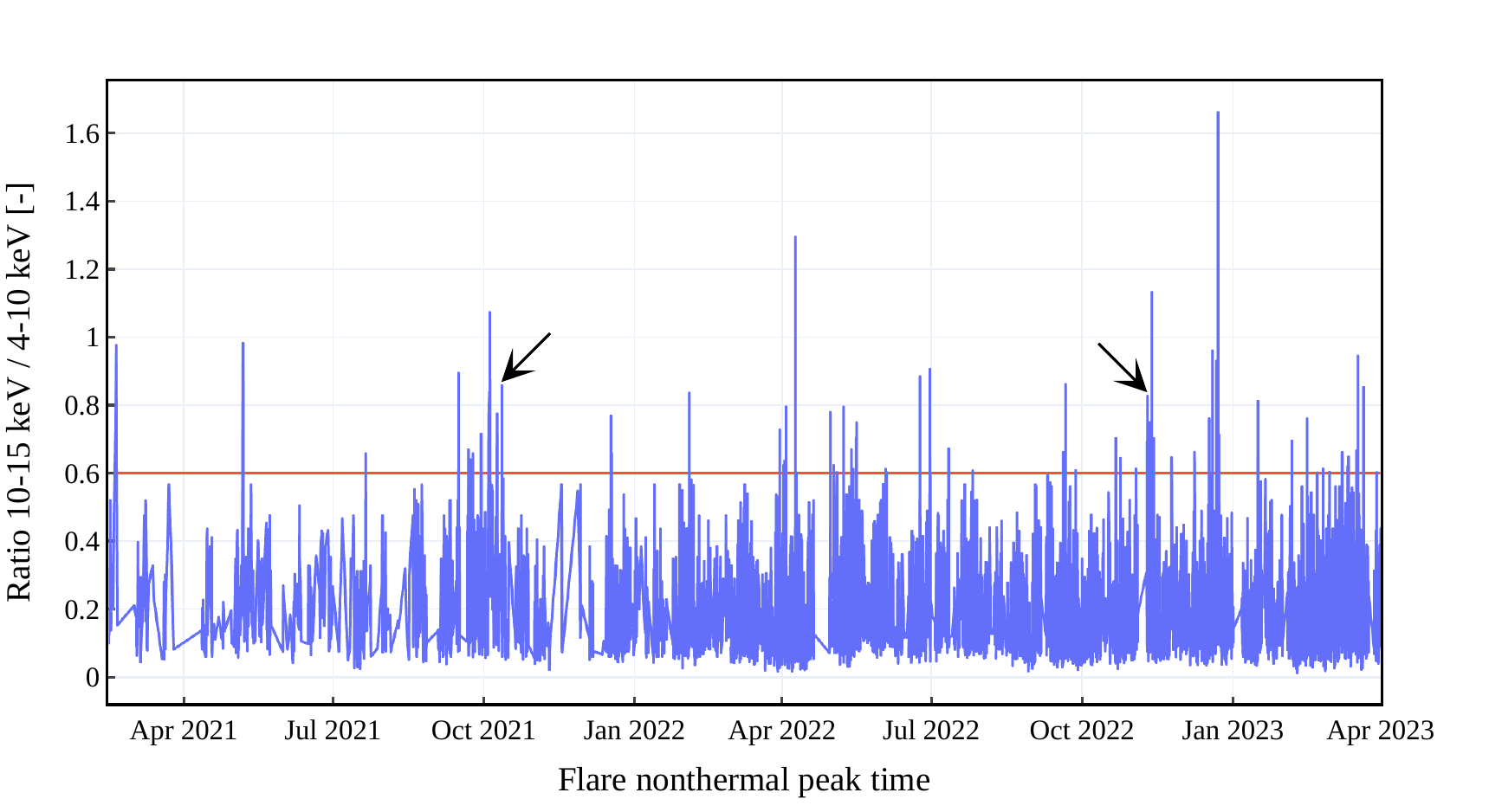}
	\caption{Ratio of STIX QL counts in the (10--15) to the (4--10)\,keV\,energy bands during the nonthermal flare peak times for the period March 2021 to April 2023. The red horizontal line indicates the 0.6 threshold. The arrows show the events under study.}
	\label{F-scatterplot1}
\end{figure}

While the QL data binned into coarse energy bands is continuously sent down from the spacecraft, the pixel data product required for reconstructing STIX images is only transmitted for selected events, thus reducing the number of flares of interest. For the initial selection, we required the ratio of the (10--15) to (4--10) keV counts of the QL channels to be smaller than 0.6. Fig. \ref{F-scatterplot1} shows this ratio for microflares during the timespan between March 2021 and April 2023. The red line indicates the 0.6 threshold. This condition was fulfilled by 47 out of 20 000 events in the STIX catalog. Requiring the selected events being observed from Earth as well as the STIX point of view further reduces the number to 21 candidates. 
\par
Among the 21 remaining candidates with a QL channel count ratio above 0.6, some are due to counts caused by particle events, others do not have enough counts to perform reliable STIX imaging and one has already been studied \citep{battaglia2023}. Two of the remaining events stand out in particular. The flares from October 11, 2021 and November 10, 2022 show unusually hard spectra and photons up to high energies during their nonthermal peaks.
\par
During the selected flares, the Solar Orbiter spacecraft was located at distances of 0.69\,AU (October 11, 2021) and 0.61\,AU (November 10, 2022) from the Sun. The light travel time differences between the spacecraft and Earth are 152.1 seconds for the event on October 11, 2021 and 187.2 seconds for November 10, 2022. These shifts are considered in the further analysis.

STIX pixel data was used for reconstructing X-ray images. For the spectral analysis, the spectrogram data product was used since it offers the highest time resolution available. The spectrogram data was further binned in time to increase the count statistics for performing spectral analysis with varying integration times. The spectral fitting was done with the OSPEX tool \citep{schwartz2002} available in SSWIDL. The functional fits were done with a combination of an isothermal component and a thick target model. For the STIX image reconstruction, the data was integrated over the entire flare duration. For thermal STIX images, the energy range was chosen to be 4--8\,keV, while the range 16--28\,keV was chosen to contain solely nonthermal counts, as supported by the spectroscopic analysis. For X-ray imaging and spectral fitting, the preflare background was subtracted{using the STIX preflare background files closest in time.
\par
For analysis of the flare morphology, EUV observations from the Atmospheric Imaging Assembly (AIA; \citealt{lemen2012}) on the Solar Dynamics Observatory (SDO) were used.
AIA data was processed with the standard SolarSoftware (SSWIDL) routines to level 1.5. Helioseismic and Magnetic Imager (HMI; \citealt{Scherrer2012}) LOS magnetograms were used to determine the magnetic flux density in the flaring region and to get insight into the magnetic field configuration for both events under study. The STIX images were re--projected to the SDO perspective for the comparison with AIA using standard SunPy routines \citep{sunpy2020}.
\par
The October 2021 event was also well observed by the Expanded Owens Valley Solar Array (EOVSA; \citealt{gary2018}), which provides microwave light curves at 1\,s time resolution, spectra at better than 40\,MHz frequency resolution, and images at up to 50 frequency bands in the range 1--18\,GHz. Two of EOVSA's 13 antennas were not in service on this date, so images were obtained from the 11 remaining antennas.  Additionally, the event occurred during a time when the Sun was behind the belt of geosynchronous satellites, which cause radio frequency interference affecting frequencies in the C (3.5--4.2\,GHz) and Ku (11.5--12.5\,GHz) bands. The microwave burst from this A9 flare reached a surprisingly high radio flux density of 81\,sfu (solar flux units; 1\,sfu = $\mathrm{10^{22}\,W\,m^{-2}\,Hz^{-1}}$) at its peak frequency of around 5\,GHz, and showed a prominent quasi-periodic pulsation with a period of 4\,s. Data were integrated over the microwave peak time range from 19:23:50-19:23:54\,UT to produce images following a self-calibration procedure using the standard CASA (Common Astronomy Software Applications) software. Images at representative frequencies 3.20, 4.82, 5.79, 6.77, 7.74, 8.72, 9.69, 10.67, and 13.27\,GHz were made, which show a nonthermal spectrum of peak brightness temperatures (66.7, 74.6, 62.5, 45.1, 28.3, 20.8, 15.7, 13.9, and 12.5\,MK, respectively) that peaks near 5\,GHz.
\par
For our inquiry into the November 17, 2006 flare published by \cite{Hannah2008MFlare}, we used white-light data from the Kanzelhöhe Observatory for Solar and Environmental Research \citep{kso2021} and a Transition Region and Coronal Explorer (TRACE; \citealt{trace1999}) 284\,\AA\, observation. The TRACE image was processed using the available SSWIDL routines and shifted by $x+3''$ and $y+7''$ to account for pointing differences, as was done in \cite{Hannah2008MFlare}. The Kanzelhöhe observation was corrected for differential rotation to match the RHESSI image, since the observation closest in time to the flare was taken around 5\,hours later. We further created a RHESSI X-ray image using the same detectors and energy range (12--60\,keV) as used in \cite{Hannah2008MFlare}.

\section{Results}
\subsection{October 11, 2021}
Fig. \ref{F-oct2021STIXGOES} shows STIX, EOVSA and GOES lightcurves for the October 11, 2021, microflare under study with X-ray and radio peak times around 19:24\,UT. After background subtraction of the preflare interval 18:28--18:42\,UT, the flare reaches GOES class A9. 
\par
Spectral fits to the STIX X-ray spectrum for selected time intervals are shown in Fig. \ref{F-Oct2021_STIXFits1}. At the first time interval integrated during the impulsive (HXR peak) phase from 19:23:48,UT--19:24:04\,UT (left panel), a hard spectrum with $\delta = (2.98 \pm 0.25)$ is observed. STIX records sufficient counts for reliable spectral analysis up to 76\,keV. The thermal plasma is described by a temperature of $T=(10.6 \pm 1.4)$\,MK and emission measure $\mathrm{EM}=(36.87 \pm 0.04)\, 10^{45}$\,$\mathrm{cm^{-3}}$. The spectrum in the right panel during the decay phase is best fitted with a single isothermal component with $T=(10.5 \pm 0.91)$\,MK and $\mathrm{EM}=(28.03 \pm 0.02) \,10^{45}$\,$\mathrm{cm^{-3}}$.
\begin{figure}[!h]   
	\includegraphics[width=0.5\textwidth,clip=]{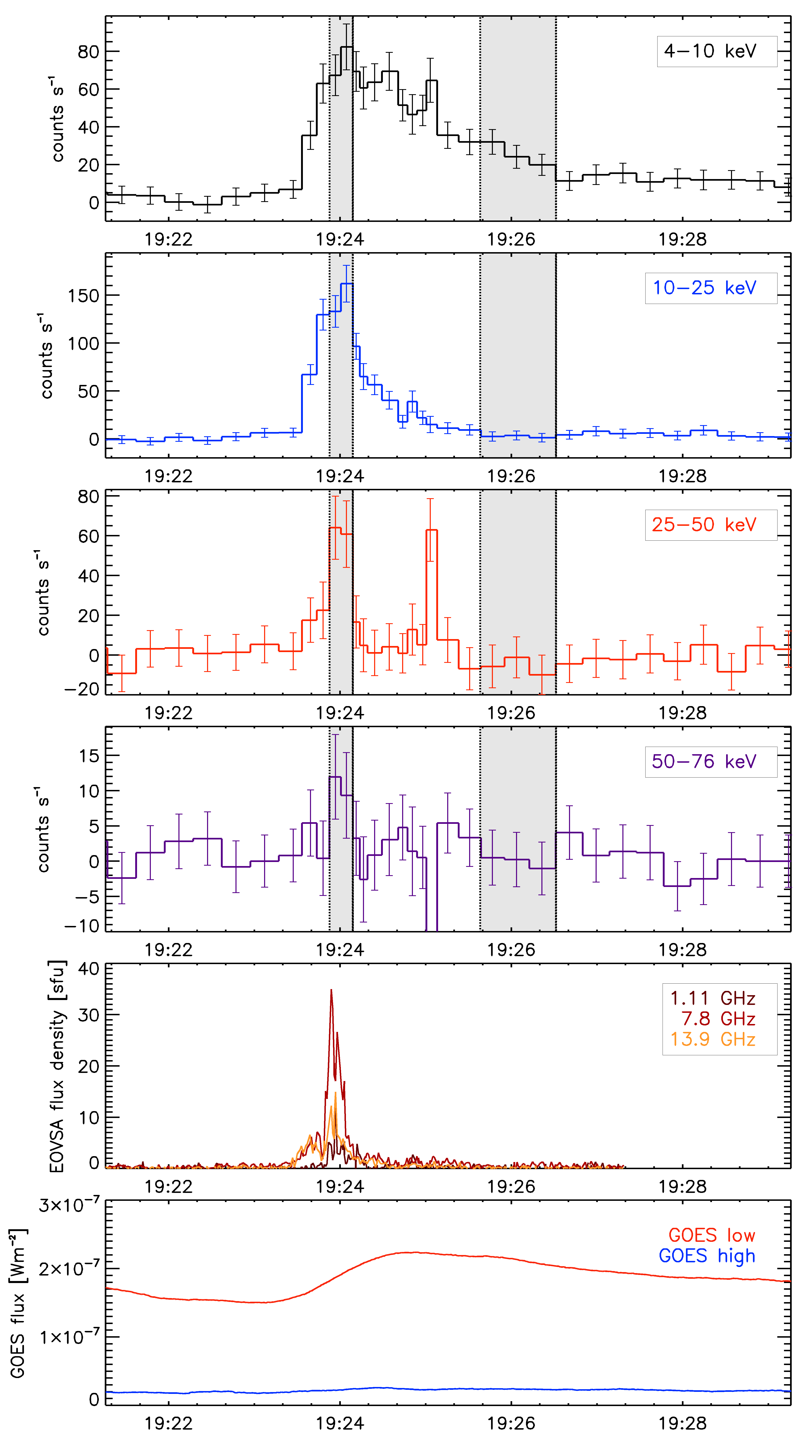}
	\caption{Lightcurves of the October 11, 2021 flare. Top four panels: STIX X-ray counts in four selected energy ranges from 4 to 76\,keV. Shaded areas indicate timespans considered in the spectral fitting. Second to last: EOVSA microwave lightcurves for three selected frequencies. Bottom: GOES 1--8\,\AA\, (red) and 0.5--4\,\AA\, (blue) soft X-ray fluxes.}
	\label{F-oct2021STIXGOES}
\end{figure} 

\begin{figure}[!h]   
	\includegraphics[width=0.4\textwidth,clip=]{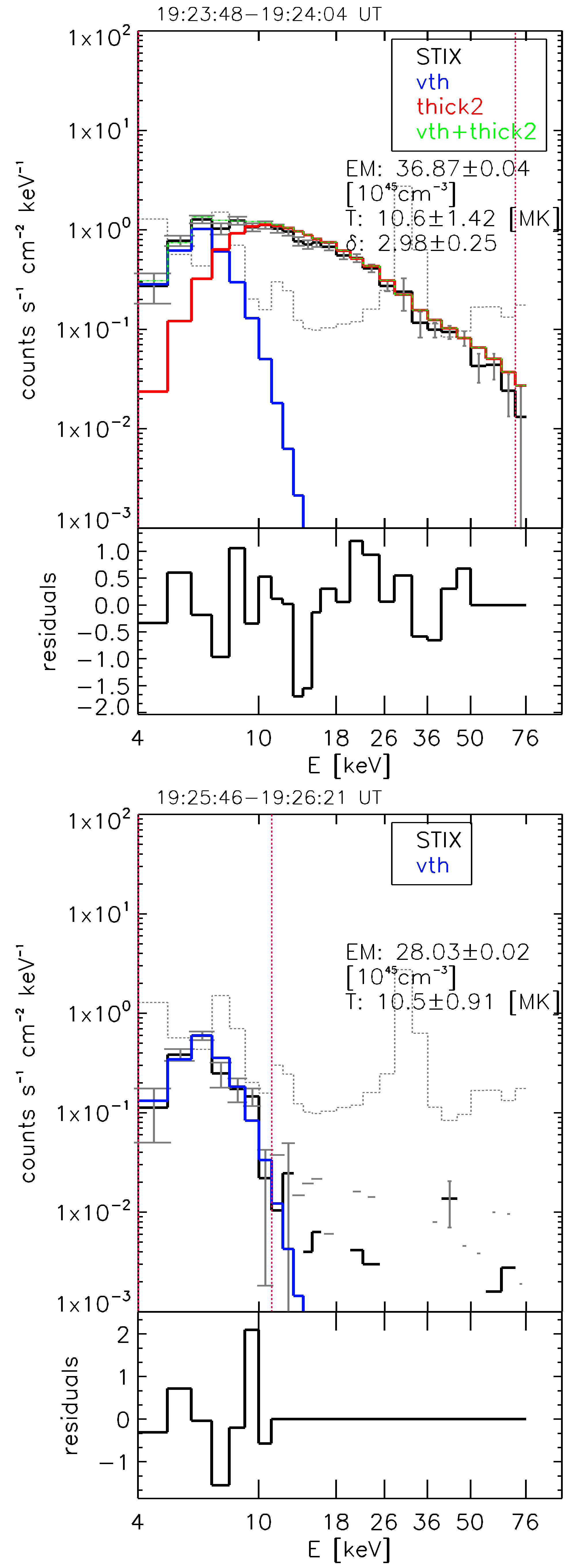}
	\caption{Background subtracted STIX count spectra (black) and best fit (green) from the sum of an isothermal (blue) and a thick target (red) component for October 11, 2021 during the peak (top panel) and decay (bottom panel) phase. Gray vertical lines on the data points indicate the error bars of the count rate. The gray dashed line shows the preflare background spectrum. The red vertical lines show the energy range considered in the fitting. Integration times for the spectral fitting are indicated by gray lines in the top panel of Fig. \ref{F-oct2021STIXGOES}.}
	\label{F-Oct2021_STIXFits1}
\end{figure} 

\par
Fig. \ref{F-131AOverview_Oct2021} (top row) shows an overview of this event in the 131\,\AA\, AIA EUV channel, which during flares is mainly sensitive to plasma around 10\,MK \citep{lemen2012}. In the leftmost panel, three regions that reveal distinct brightening during the flare are observed. The middle panel at 19:24\,UT shows loops of hot plasma connecting the flare kernels from the previous panel and an additional loop in the north as well as a jet--like feature in the south. In the right panel taken at 19:31\,UT, i.e. after the flare energy release has seized, the emission from hot plasma sampled by the 131\,\AA\, filter continually decreases.
The bottom row of Fig. \ref{F-131AOverview_Oct2021} shows the corresponding evolution in the chromospheric AIA 1600\,\AA\, filter with three (left) and four (middle panel) distinct footpoint brightenings being visible corresponding to the brightenings and loops observed in the 131\,\AA\, images in the top row. Notably, one of the AIA 1600\,\AA\, brightenings is observed inside the sunspot.
\par
The evolution observed in the AIA 1600\,\AA\,filter is further elaborated in Fig. \ref{F-1600AOverview_Oct2021} (see also the accompanying movie) which shows a preflare AIA 1600\,\AA\, image from 19:20\,UT as a reference in panel a). Panel b) reveals four areas of increased brightness during the flare impulsive phase at 19:23:50\,UT. The brightening to the far west is located in the umbra of the sunspot. In the base difference image shown in panel d), these enhancements are seen more clearly. Overplotted are contours of STIX spectral images reconstructed in the 4--8\,keV (red) and 16--28\,keV (green) energy bins. The same contours are also shown on top of an HMI LOS magnetogram in panel c). The nonthermal STIX image recovers two flare kernels visible in the AIA 1600\,\AA\, images. The thermal source seen by STIX is located between two flare kernels, namely the most western kernel that is inside the sunspot umbra (located in negative magnetic polarity) and the one closely outside the sunspot penumbra to the east (located in positive polarity). The two other kernels are located in negative polarity regions of the trailing plage region.
\par
The STIX total HXR count fluxes in the 16--28\,keV energy range in the footpoints are 0.05 for the umbral kernel and 0.02\,[$\mathrm{cnts\,s^{-1}cm^{-2}arcsec^{-2}keV^{-1}}$]\,for the eastern kernel. The mean magnetic flux density obtained from the HMI LOS magnetogram in the footpoints observed in 1600\,\AA\, is 1544\,G with a standard deviation of $\pm$354 for the western footpoint within the sunspot, ($-95\pm107$)\,G  for the northernmost footpoint, ($-96\pm122$)\,G for the north-east and ($-37\pm52$)\,G\,in the southern location.
\par
Fig. \ref{F-EOVSAImages} shows contours of EOVSA radio images at different frequencies plotted on top of AIA 131\,\AA\, and 1600\,\AA\, filtergrams. In the left panel 20, 50 and 70\,\% contours are shown to illustrate the lower frequency sources extending along the flare loops to the north. In the right panel, only 50 and 70\,\% levels are shown. This panel clearly shows that the center of the source location gets shifted toward the western flare kernel with increasing frequency, indicative of gyrosynchrotron emission from low in the chromosphere.

\begin{figure}[!h]   
	\includegraphics[width=0.5\textwidth,clip=]{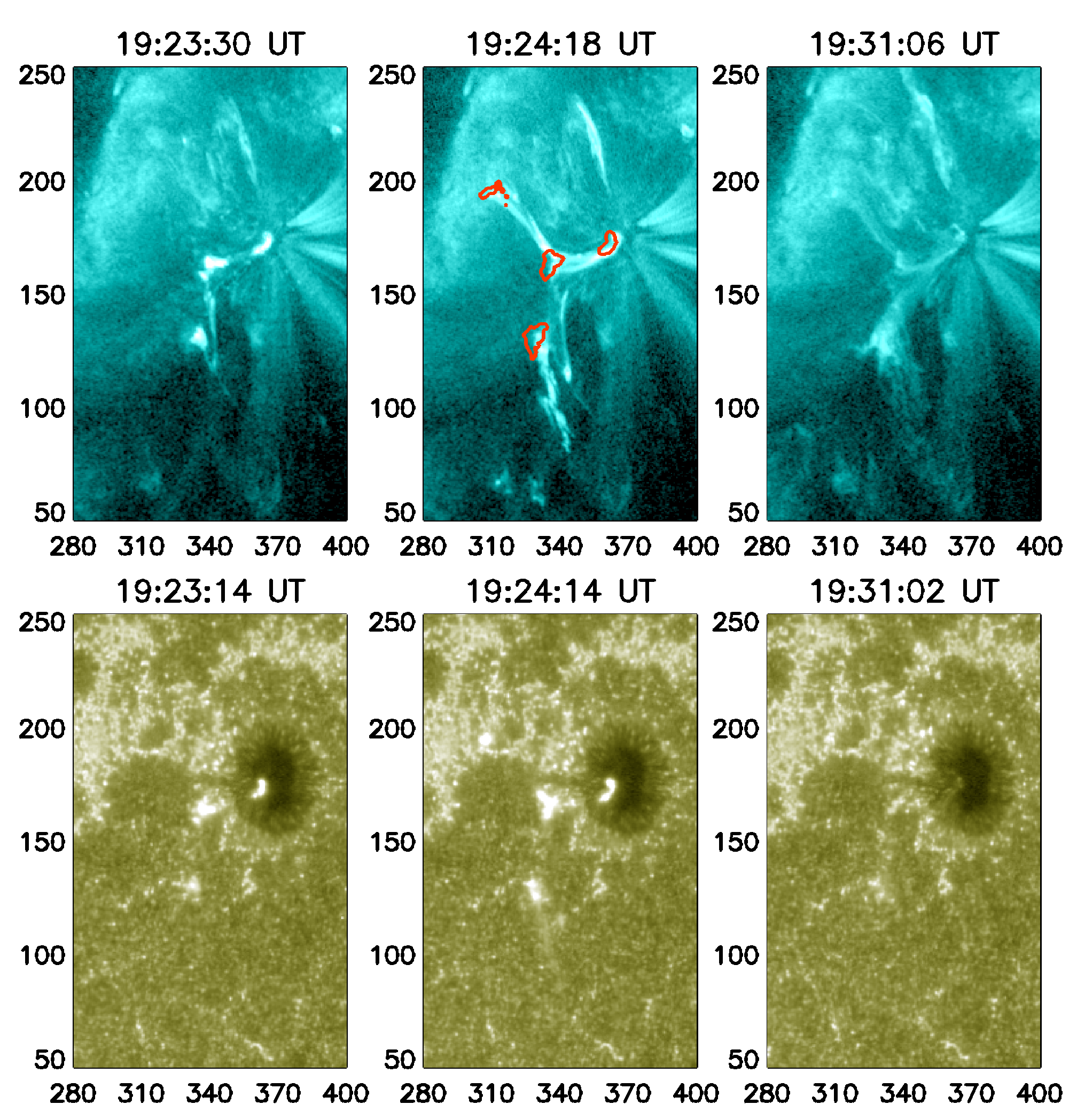}
	\caption{Top: AIA 131\,\AA\, filtergrams over the course of the flare on October 11, 2021. Bottom: AIA 1600\,\AA filtergrams. Red contours in the top middle panel show footpoint locations derived from the AIA 1600\,\AA \,base difference image shown in Fig. \ref{F-1600AOverview_Oct2021}.
    Units are given in arcseconds. An animated version of this figure is included in the online supplementary electronic material. \href{https://drive.google.com/file/d/1W9iOypk_ZF0gdWQhvyWTRpEvGNQaoR5l/view?usp=sharing}{[Google Drive]}}
	\label{F-131AOverview_Oct2021}
\end{figure}

\begin{figure}[!h]   
	\includegraphics[width=0.5\textwidth,clip=]{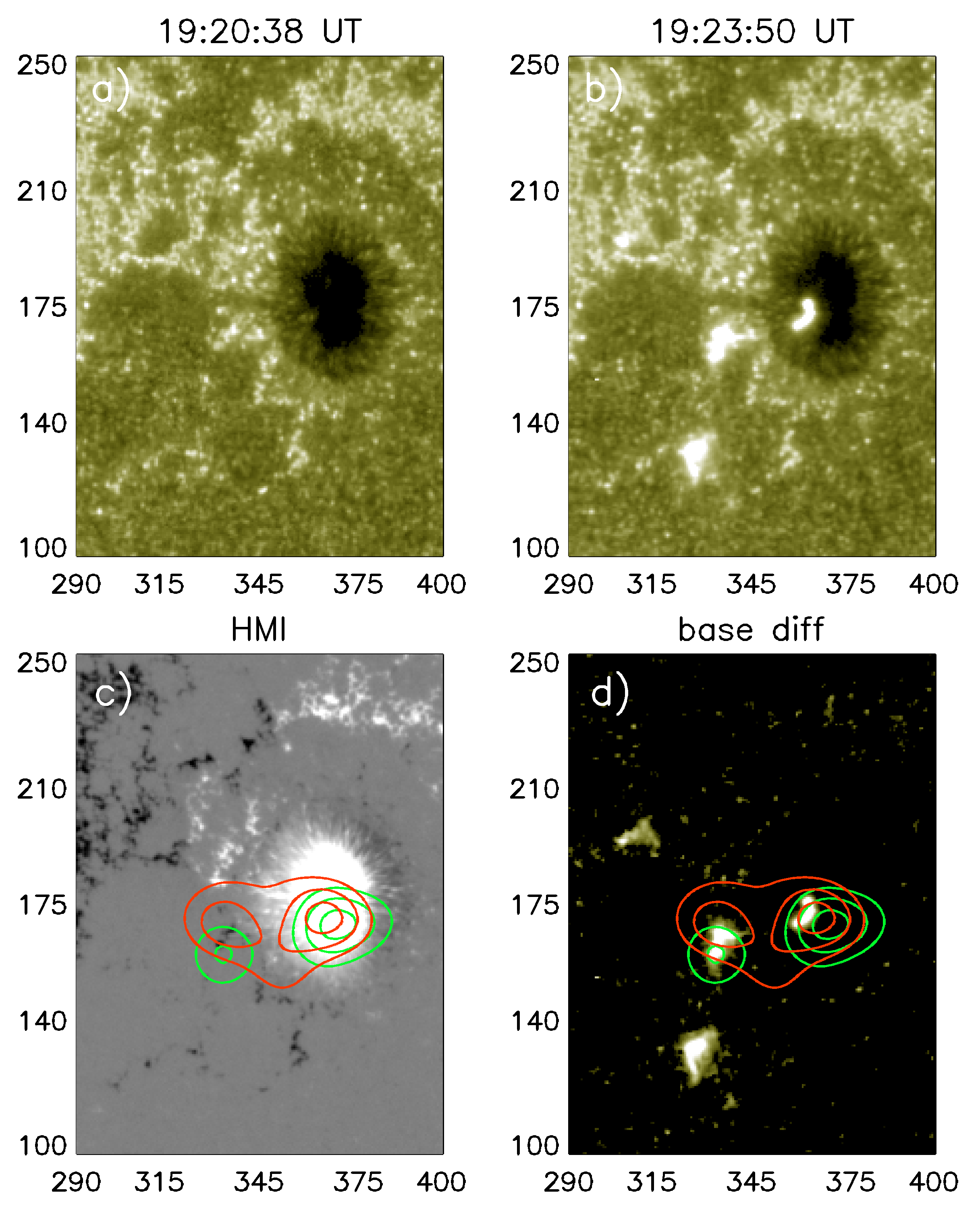}
	\caption{AIA 1600\,\AA\, images (panels a, b), HMI LOS magnetogram (c) and 1600\,\AA\,difference image (d) for October 11, 2021. Green contours show the STIX image in the 16--28\,keV energy range, red in 4--8\,keV (30, 60, 90 \% of the maximum intensity). The HMI image is scaled from $-$700 to $+$700\,G. Units are given in arcseconds. Integration times are 19:23:21--19:24:36\,UT for the nonthermal and 19:24:51--19:26:31\,UT for the thermal image respectively.}
	\label{F-1600AOverview_Oct2021}
\end{figure}

\begin{figure}[!h]   
	\includegraphics[width=0.5\textwidth,clip=]{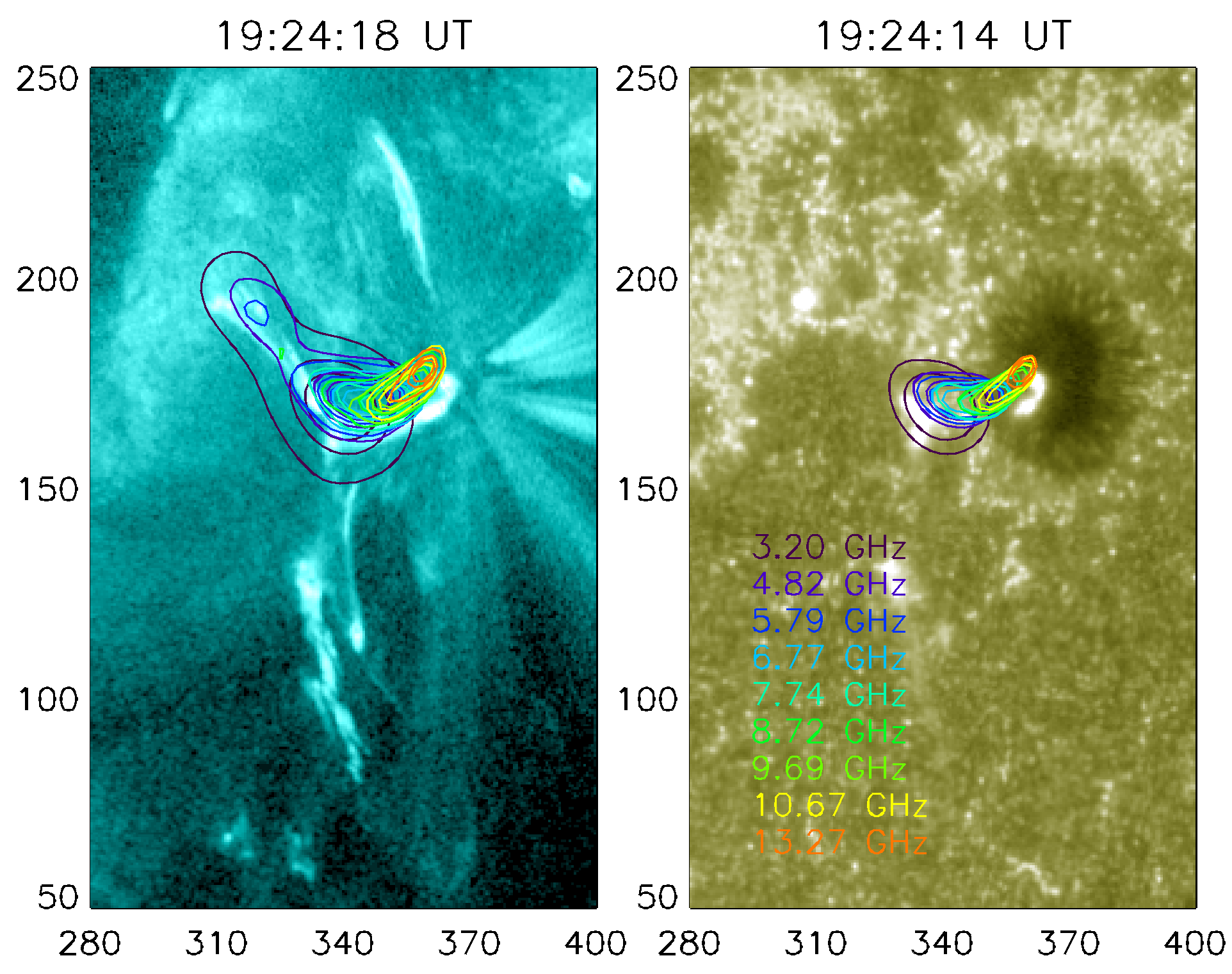}
	\caption{Contours of EOVSA radio images at various frequencies from 3.20 to 13.27\,GHz plotted on top of an AIA 131\,\AA\,(left) and 1600\,\AA\,(right) filtergram for the event on October 11, 2021. The color code for the different EOVSA frequencies is the same in both images. The contours in the left panel are 20, 50 and 70\,\% of the maximum intensity; in the right panel 50 and 70\%.}
	\label{F-EOVSAImages}
\end{figure} 

\subsection{November 10, 2022}
\begin{figure}[!h]   
	\includegraphics[width=0.5\textwidth,clip=]{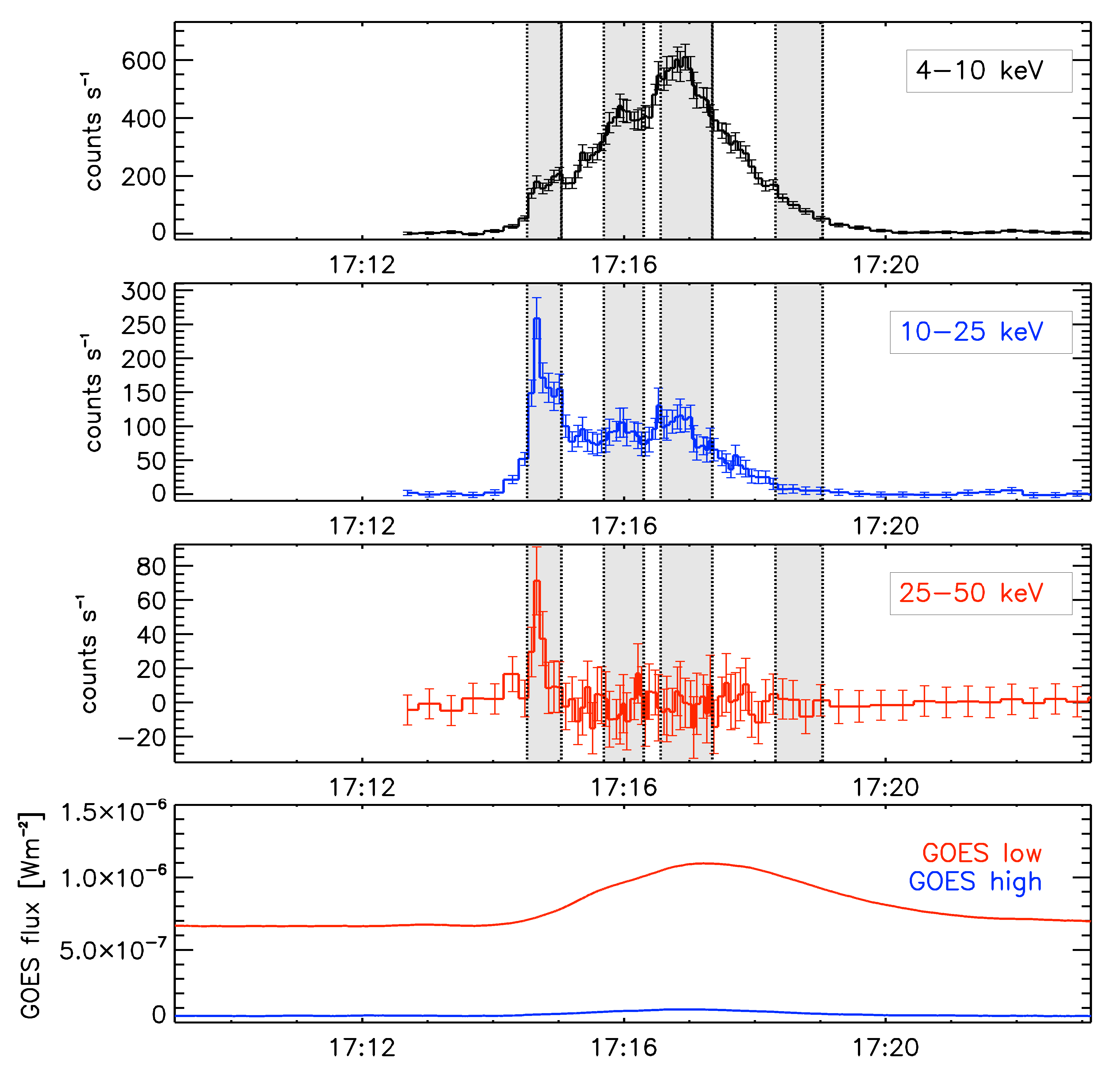}
	\caption{Lightcurves of the November 10, 2022 flare. Top panels: STIX counts in selected energy bins from 4 to 50 keV. Bottom: GOES 1--8\,\AA\, (red) and 0.5--4\,\AA\, (blue) soft X-ray fluxes. Shaded areas indicate timespans considered in the spectral fitting.}
	\label{F-nov2022STIXGOES}
\end{figure}

Fig. \ref{F-nov2022STIXGOES} shows STIX and GOES lightcurves for the event of November 10, 2022. The GOES flux after background subtractionof the preflare interval 17:00--17:11\,UT reaches class C1.
\par
Functional fits to the X-ray spectrum during the flare are shown in Fig. \ref{F-Nov2022_STIXFits1}. The top left panel shows the spectrum during the start of the impulsive flare. The electron distribution power--law index is $\delta=(4.08 \pm 0.23)$. Initially, the thermal plasma is fitted by a temperature of $T=(14.8 \pm 6.97$)\,MK and $\mathrm{EM}=(8.66 \pm 0.02)$\,$10^{45}$\,$\mathrm{cm^{-3}}$. The spectrum steepens to $\delta=(5.64 \pm 0.88)$ and the EM of the best isothermal fit increases to $\mathrm{EM}=(43.86 \pm 0.04) 10^{45}$\,$\mathrm{cm^{-3}}$ with a temperature of $T=(14.1 \pm 2.07)$\,MK for the first thermal peak (top right). For the main thermal peak from 17:16:22--17:17:09\,UT (bottom left panel) $\delta=(5.97 \pm 0.26)$ for the thick target and $\mathrm{EM}=(80.67 \pm 0.04) 10^{45}$\,$\mathrm{cm^{-3}}$ and $T=(13.4 \pm 1.02)$\,MK for the isothermal component. During the decay phase shown in the bottom right panel, the best fit is achieved with solely an isothermal component of $\mathrm{EM}=(187.20 \pm 0.04)\, 10^{45}$\,$\mathrm{cm^{-3}}$ and $T=(9.4 \pm 0.27)$\,MK. 
\par
Fig. \ref{F-131AOverview_Nov2022} shows the evolution of the flare in the AIA 131\,\AA\, filter (see also the accompanying movie). Before the main flare is observed in the STIX and GOES lightcurves (Fig. \ref{F-nov2022STIXGOES}), a small loop system brightens up in the first frame at 17:12\,UT. The main event which coincides with the rise in nonthermal and thermal emission shows a single loop which is clearly enhanced in the second frame at 17:15\,UT. This enhancement is accompanied by a southward jet which is still active at 17:17\,UT. In the bottom right frame at 17:21\,UT the postflare loop is observed.  
\par
The chromospheric response and the photospheric magnetic field are shown in Fig. \ref{F-1600AOverview_Nov2022}. In the AIA 1600\,\AA\, difference image (panel d) between the frames shown in panels a) and b), three areas of increased brightness are observed. Contours of these areas are overplotted in blue and purple in panel c). The blue contours show footpoints covering opposite magnetic polarities, with the positive footpoint being located within the sunspot penumbra. The contour shown in purple covers a region of mixed magnetic polarity which indicates the remnant of the separate smaller loop system which brightened up around 17:12\,UT.
 
Contours of the reconstructed STIX image are overplotted in panels c) and d). The nonthermal 16--28\,keV image (green) shows two sources which coincide with the chromospheric response to the flare electrons observed in the 1600\,\AA\, filter. In the thermal 4--8\,keV range (red), a single extended source between the nonthermal footpoints is observed. The STIX contours over the HMI LOS magnetogram in panel c) show that the northern HXR footpoint is located within the sunspot penumbra. The mean magnetic LOS flux density within the AIA 1600\,\AA\, flare kernels are (1568$\pm$350)\,G  for the northern and ($-$117$\pm$137)\,G for the southern kernel.
\par
The STIX summed HXR count fluxes in the 16--28\,keV energy range are 0.7 for the umbral kernel and 0.6\,[$\mathrm{cnts\,s^{-1}cm^{-2}arcsec^{-2}keV^{-1}}$]\,in the southern footpoint outside the sunspot.
\par
\begin{figure}[!h]   
	\includegraphics[width=0.48\textwidth,clip=]{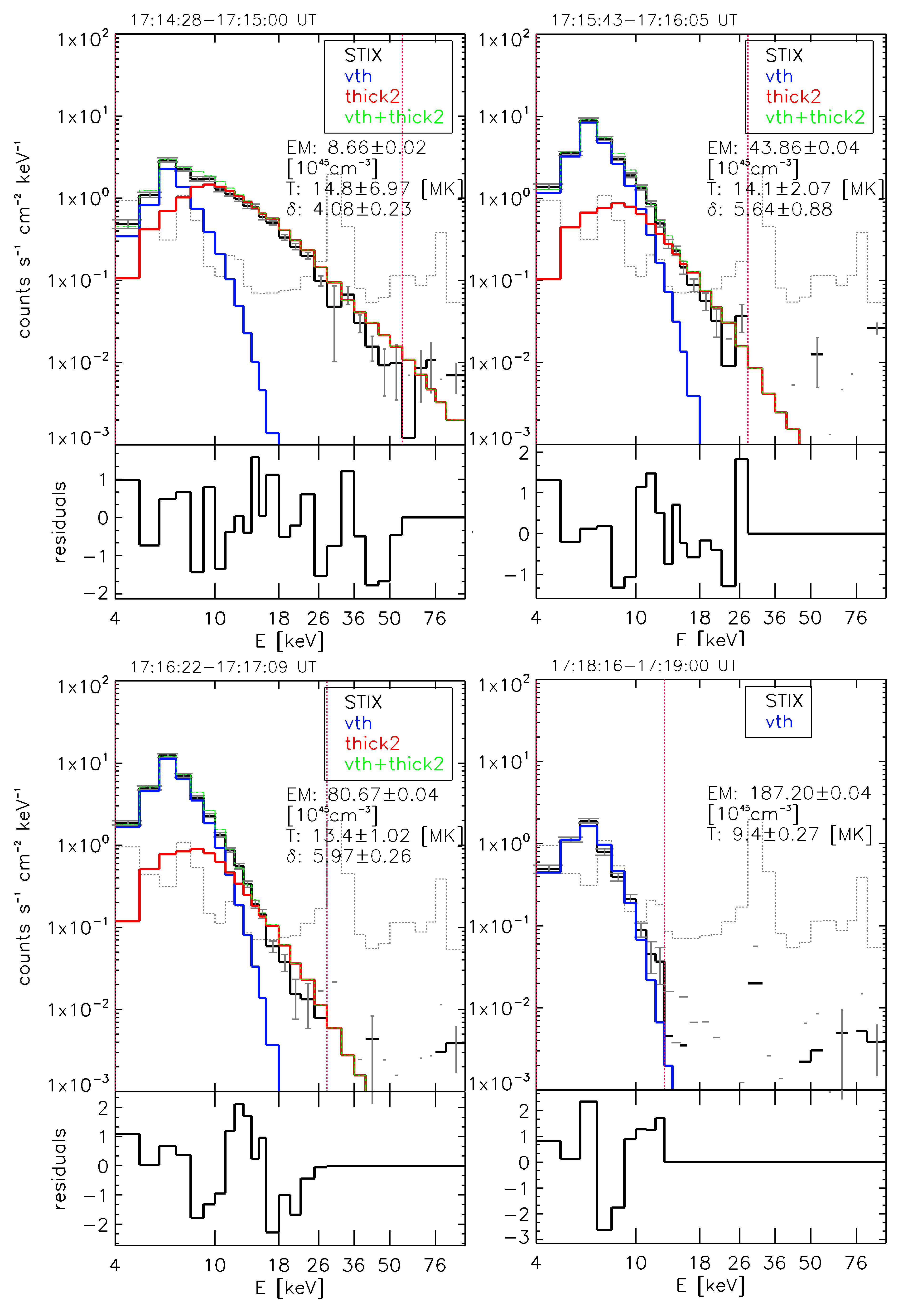}
	\caption{Background subtracted STIX count spectra (black) and best fit (green) from the sum of an isothermal (blue) and a thick target (red) component for November 10, 2022 for four time intervals during the flare evolution (indicated in Fig. \ref{F-nov2022STIXGOES}). Gray vertical lines on the data points indicate the error bars of the count rate. The gray dashed line shows the preflare background spectrum. The red vertical lines show the considered energy range.}
	\label{F-Nov2022_STIXFits1}
\end{figure} 

\begin{figure}[!h]   
	\includegraphics[width=0.5\textwidth,clip=]{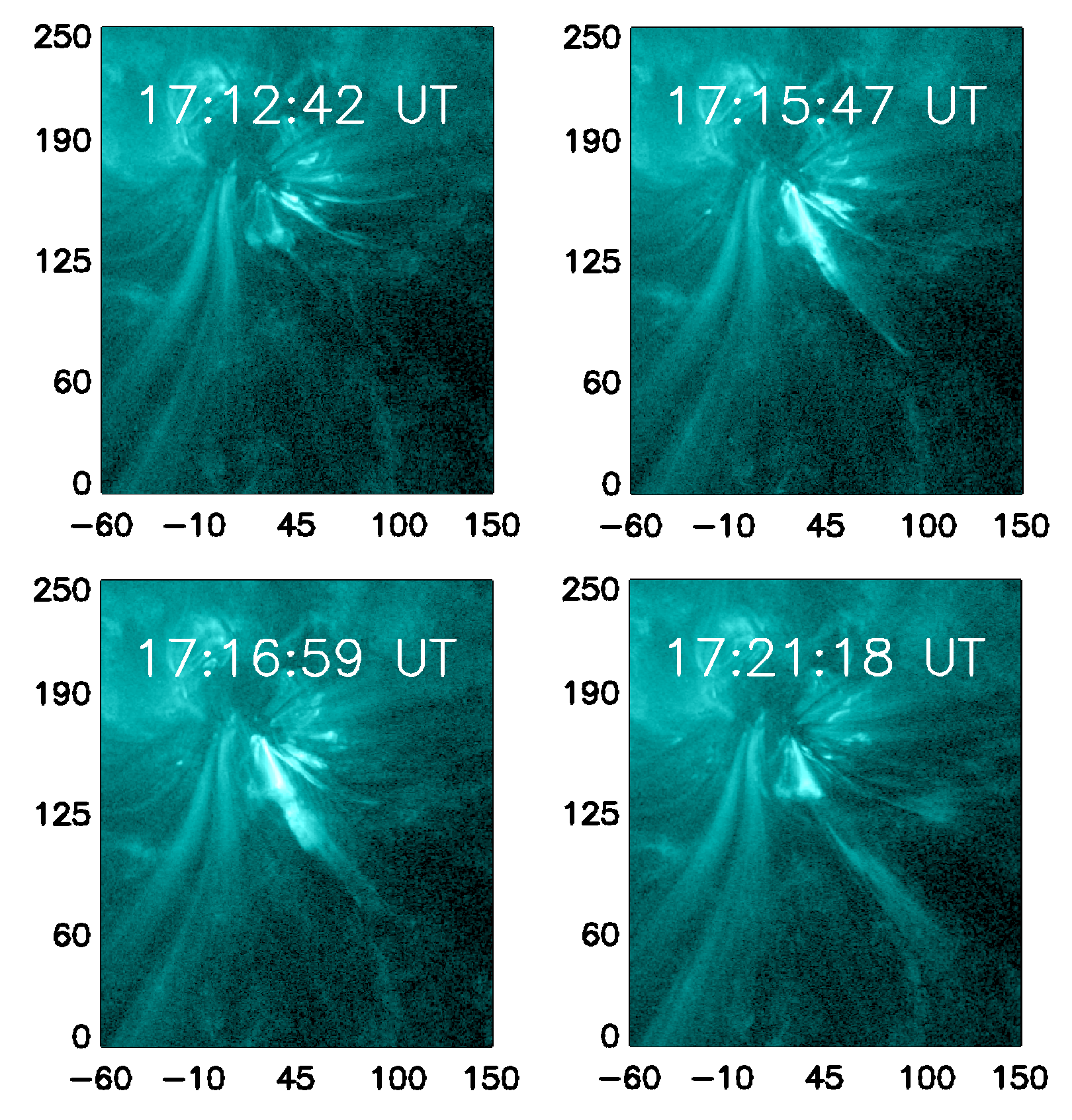}
	\caption{AIA 131\,\AA\, filtergrams over the course of the flare on November 10, 2022. Units are given in arcseconds. An animated version of this figure is included in the online supplementary electronic material. \href{https://drive.google.com/file/d/1ginhs0IuskbyDGrNR_WsjdpT6WisQbFP/view?usp=sharing}{[Google Drive]}}
	\label{F-131AOverview_Nov2022}
\end{figure} 

\begin{figure}[!h]   
	\includegraphics[width=0.5\textwidth,clip=]{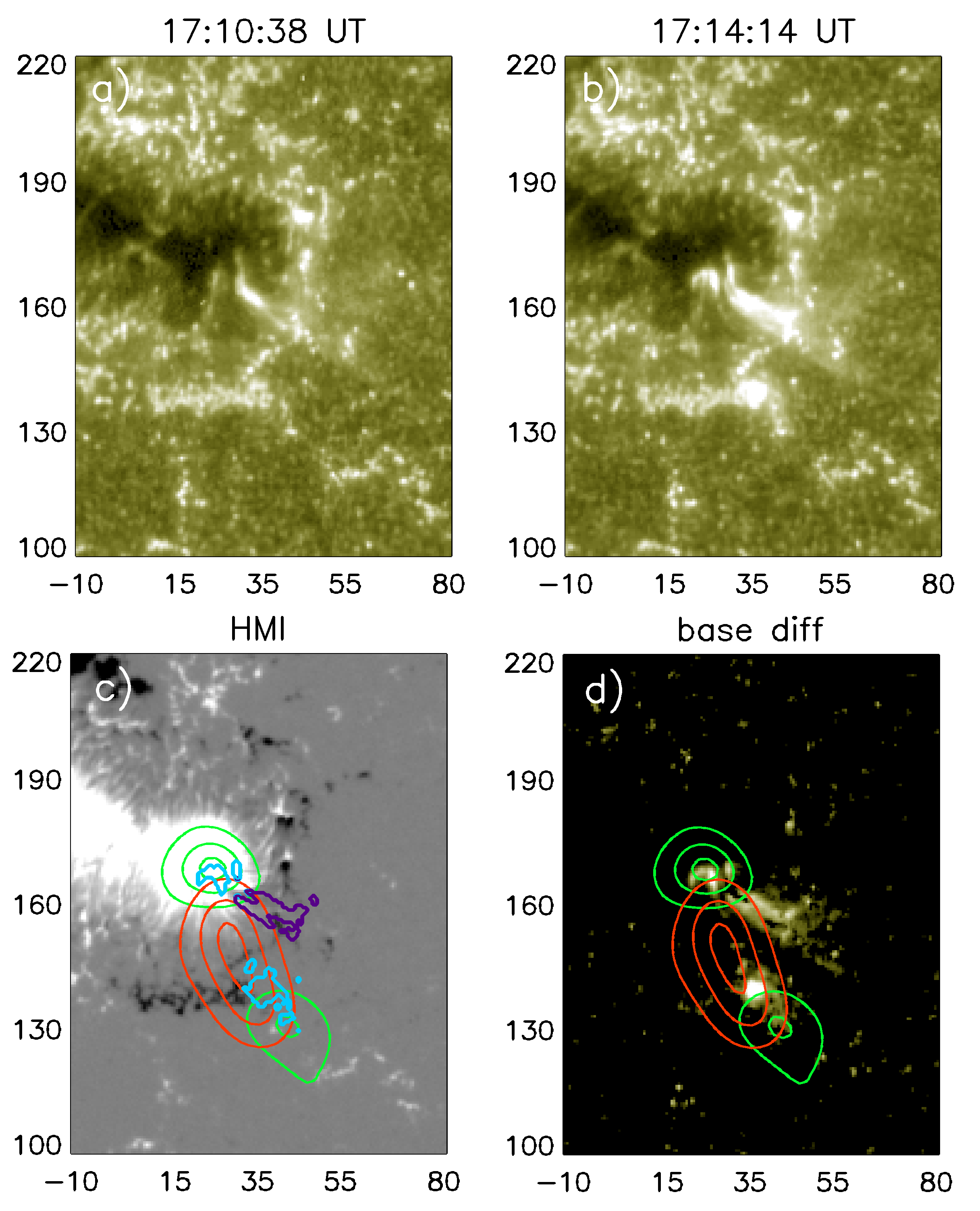}
	\caption{AIA 1600\,\AA\, images (panels a, b), difference image between them (d) and HMI LOS magnetogram (c) for the November 10, 2022 flare. Green contours show the STIX image in the 16--28\,keV energy range, red in 4--8\,keV (30, 60, 90 \% of the maximum intensity). The HMI image is scaled from $-$1000 to $+$1000\,G. Blue contours in the HMI image show the areas used for the calculation of the mean magnetic flux densities. The purple contour indicates the area from the AIA 1600\,\AA\,base difference image considered to be part of a separate, smaller loop system. Units are given in arcseconds. Integration times are 17:14:05--17:15:25\,UT for the nonthermal and 17:15:25--17:17:25\,UT for the thermal image respectively.}
	\label{F-1600AOverview_Nov2022}
\end{figure} 

\section{Discussion}
We studied two microflares of GOES classes A9 and C1 with unusually hard X-ray spectra and strong nonthermal emission up to high energies during the impulsive phase observed by STIX.
\par
The photon power law indices assuming the relation for thick--target bremsstrahlung $\gamma = \delta-1$, during the flare impulsive phase are $\gamma=1.98$ for the October 11, 2021 event and $\gamma=3.08$ in the November 10, 2022 flare. Both events show photons up to high energies considering their small GOES classes, with photon energies up to 76 and 50\,keV for the October and November flare respectively. The spectrum of the previously hardest microflare studied in \cite{Hannah2008MFlare} was best fitted by $\gamma=2.4$ and showed photon energies up to about 50\,keV. 
\par
For both events in this study, one of the flare footpoints was located directly within a sunspot during the onset, rather than moving there over the course of the flare. In the October 11, 2021 A9 flare, one footpoint is located within the sunspot umbra, in strong fields with a mean flux density of (1544$\pm$354)\,G below the AIA 1600\,\AA\,kernel. In the flare of November 10, 2022, the northern flare kernel partly covers the sunspot penumbra and umbra with a mean magnetic LOS flux density of (1568$\pm$350)\,G. 
These magnetic flux densities are significantly higher than typical mean values for the mean flux densities in flare ribbons reported by \cite{kazachenko2017} who did not report values above 1000\,G with a 20th--80th percentile range of 408--675\,G and the range of 100--800\,G reported in \cite{tschernitz2018} even though they considered flares up to GOES class X17. From extrapolating the relation between flare GOES class and mean magnetic flux density in the flare ribbons of eruptive flares published by \cite{tschernitz2018} to smaller events, expected values of the flux densities for the A9 and C1 flares in our study are 30 and 67\,G, which is one to two orders of magnitude lower than the observed values.
\par
Flares with differences in the magnetic field strength of the flare loop footpoints often show an asymmetry in the HXR fluxes, with the stronger HXR emission being located at the site of weaker magnetic field due to asymmetric magnetic mirroring (\citealt{aschwanden1999,yang2012}). Despite the significantly stronger magnetic fields in the footpoints located in the sunspots, the STIX flares under study do not show stronger HXR emission from the footpoints located at the weaker magnetic fields outside the sunspot. Possible reasons for such deviations are the electron injection site being located closer to the brighter footpoint \citep{goff2004,falewicz2007}, varying plasma density along the flare loop \citep{falewicz2007} or differences in the magnetic field convergence along both directions \citep{yang2012}. 
\par
From EOVSA microwave images of the October 11, 2021 event, we find that the locations of the radio sources at typical gyrosynchrotron frequencies shift towards the umbral flare kernel with increasing frequency, in agreement with modelling results of radio emission from an asymmetric loop \citep{bastian2000}. Microwaves are typically produced by relatively high-energy electrons (\char`\~300\,keV). The strong microwave emission seen by EOVSA for the October 2021 A9 event thus indicates that the relatively flat spectrum measured by STIX extends at least to energies of that order. The fact that the EOVSA high-frequency source is located above the sunspot umbra means that the radio-emitting electrons must be able to reach the strong umbral magnetic field region despite a relatively high mirror ratio from this asymmetric loop. 
\begin{figure}[!h]   
	\includegraphics[width=0.5\textwidth,clip=]{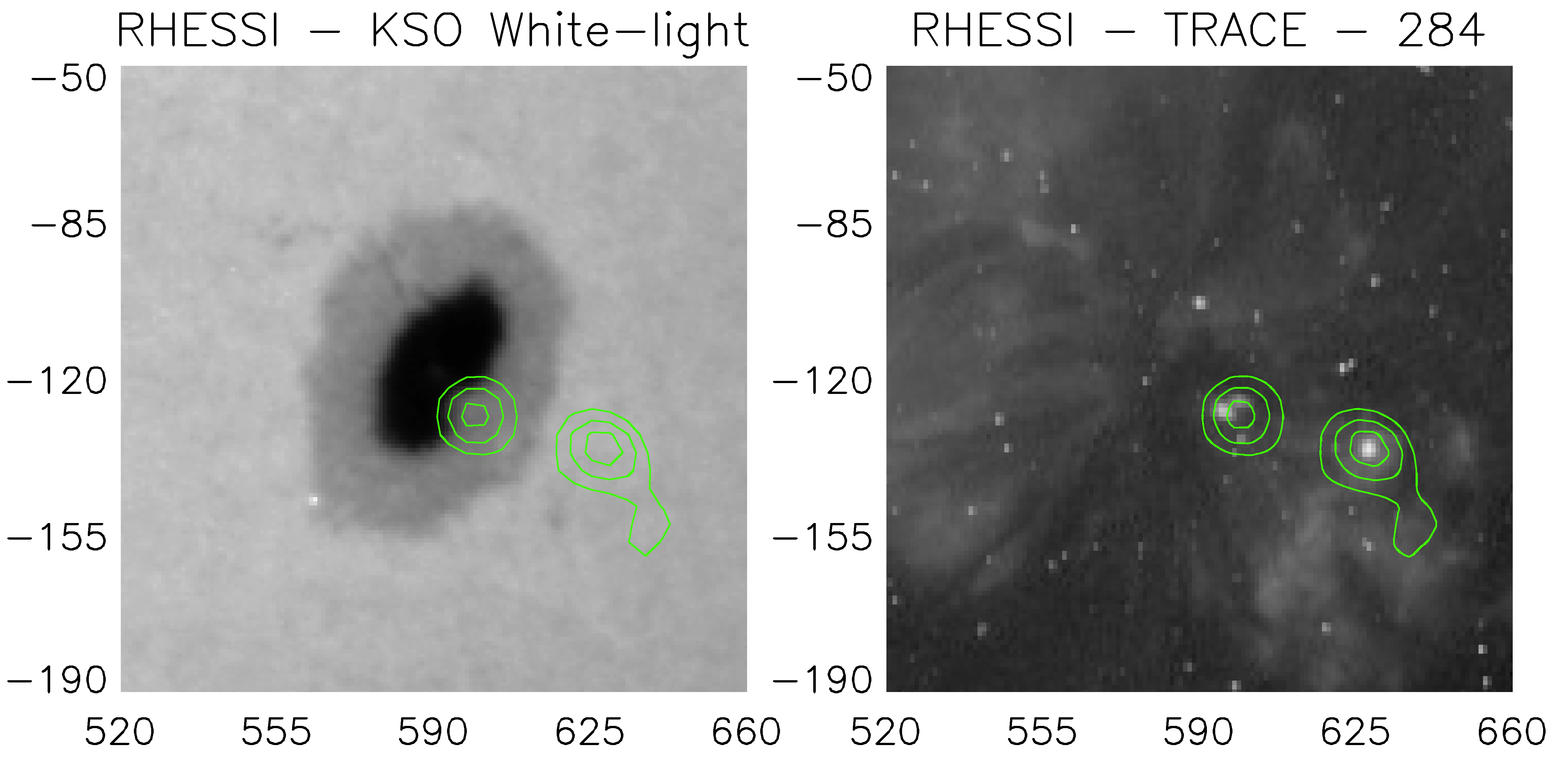}
	\caption{Left: Kanzelhöhe white-light image from November 17, 2006 and RHESSI 12--60\,keV image contours from 05:13:40--05:13:52\,in green (60, 75, 90\% of the maximum intensity). Right: TRACE 284\,\AA\, EUV filtergram from 05:14:01\,UT and same RHESSI contours in green. Units are given in arcsec.}
	\label{F-kso_rhessi}
\end{figure}
\par
In light of our findings for the two hard STIX microflares, we revisited the hard RHESSI GOES A7 microflare previously reported in \cite{Hannah2008MFlare} using Kanzelhöhe white-light and TRACE EUV data and found that the eastern contours of the nonthermal (12--60\,keV) RHESSI image integrated over the impulsive peak are located within the sunspot penumbra and part of the umbra (Fig. \ref{F-kso_rhessi}), analogous to the analyzed STIX events. Umbral flares are rare occurrences as flares usually occur within active regions but not directly in sunspots, but cases have been reported (e.g., \cite{tang1978}, \cite{joshi1992} and \cite{li1998}). However, these are all studies of regular flares. So far, there are no studies of umbral microflares. In this study, we find that in the two STIX and the one RHESSI microflare with very hard X-ray spectra, at least one footpoint is located directly in a sunspot. We therefore conclude that the characteristics of the strong photospheric magnetic fields inside sunspot umbrae and penumbrae where the flare loops are rooted play an important role in the generation of the exceptionally hard X-ray spectra in these microflares.

\par

 \begin{acknowledgements}
JS, AMV and ECD acknowledge the Austrian Science Fund (FWF): I4555-N. AFB is supported by the Swiss National Science Foundation Grant 200021L\_189180 for STIX. Solar Orbiter is a space mission of international collaboration between ESA and NASA, operated by ESA. The STIX instrument is an international collaboration between Switzerland, Poland, France, Czech Republic, Germany, Austria, Ireland, and Italy. EOVSA is supported by NSF grant AGS-2130832 to New Jersey Institute of Technology. DG acknowledges support from NASA grant 80NSSC18K1128. SDO image data courtesy of NASA/SDO and the AIA science team. HMI data courtesy of NASA/HMI and the HMI science team.
\end{acknowledgements}

%
%

\bibliographystyle{aa} 
\bibliography{biblio.bib} 

\end{document}